# Metamagnetic and electronic transitions in charge-ordered $Nd_{0.50}Ca_{0.47}Ba_{0.03}MnO_3$ manganite

K. R. Mavani[†], P. L. Paulose*

*Department of Condensed Matter Physics and Materials Science, Tata Institute of Fundamental Research, Colaba, Mumbai 400 005, India*

## ABSTRACT

The $ABO_3$ type charge-ordered antiferromagnetic $Nd_{0.50}Ca_{0.50}MnO_3$ (NCMO) manganite is doped at $A$-site by 3 % of $Ba^{2+}$ for $Ca^{2+}$. The resulting system, $Nd_{0.50}Ca_{0.47}Ba_{0.03}MnO_3$ (NCBMO), is studied for the effects of Ba doping on the magnetic and electronic properties. On application of magnetic field to NCBMO, strongly correlated successive sharp metamagnetic and electronic transitions are observed from antiferromagnetic-insulating to ferromagnetic-metallic state at 2.5 K. The critical magnetic field ($H_c$) required for metamagnetism is found to reduce drastically from 15 T for undoped NCMO to 3 T for NCBMO. On increasing the temperature, the $H_c$ of NCBMO passes through a minimum. This behavior of $H_c$ of NCBMO contrasts to that of NCMO. The results are discussed in context of $A$-site cation disorder and size.

*\* E-mail: paulose@tifr.res.in.*

[†]*Present address:*
*Institute of Laser Engineering, Osaka University, Osaka, Japan.*
*E-mail: mavani-k@ile.osaka–u.ac.jp; krushna1@gmail.com.*

The $ABO_3$ type manganites show strongly correlated magnetic and electronic properties [1]. Half-doped manganites, (with 1:1 ratio of $Mn^{3+}$ and $Mn^{4+}$) such as $Nd_{0.50}Ca_{0.50}MnO_3$ (NCMO), display a charge-exchange (CE) type charge-ordered insulating state. On application of high magnetic field of 15 T at a low temperature of ~4 K, NCMO single crystal shows melting of charge-ordered state with a metamagnetic transition to ferromagnetic-metallic state [2]. Earlier studies show that disorder can largely affect the magnetic and electronic properties of manganites, affect the CE type charge-ordered state [3] and may also reduce the critical magnetic field ($H_c$) required for a metamagnetism in such manganites [4, 5]. With an objective to study the disorder effects on correlated magnetic and electronic properties of NCMO system, we have doped 3% $Ba^{2+}$ (1.47 Å) for $Ca^{2+}$ (1.18 Å) in NCMO to induce $A$-site cation size disorder ($\sigma^2 = \sum x_i r_i - <r_A>^2$). We report here the drastically modified magnetic and electronic properties of $Nd_{0.50}Ca_{0.47}Ba_{0.03}MnO_3$ (NCBMO) by this fractional doping of Ba.

The compound of $Nd_{0.50}Ca_{0.47}Ba_{0.03}MnO_3$ was synthesized by the standard solid-state reaction, as described in ref. 4. The Rietveld analysis (by FULLPROF program code) of X-ray powder diffraction data shows that the compound is single phase and forms in orthorhombic symmetry (space group: *Pnma*, No. 62). The magnetization and resistivity measurements were carried using SQUID (Quantum Design), VSM (Oxford) and PPMS (Qunatum Design).

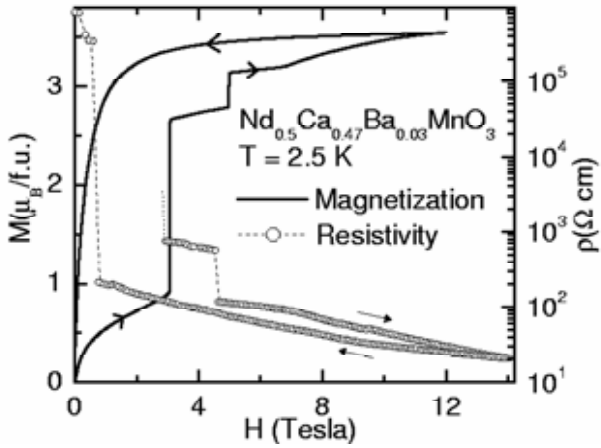

*Fig. 1: Magnetization and resistivity isotherms for $Nd_{0.50}Ca_{0.47}Ba_{0.03}MnO_3$ compound at 2.5 K.*

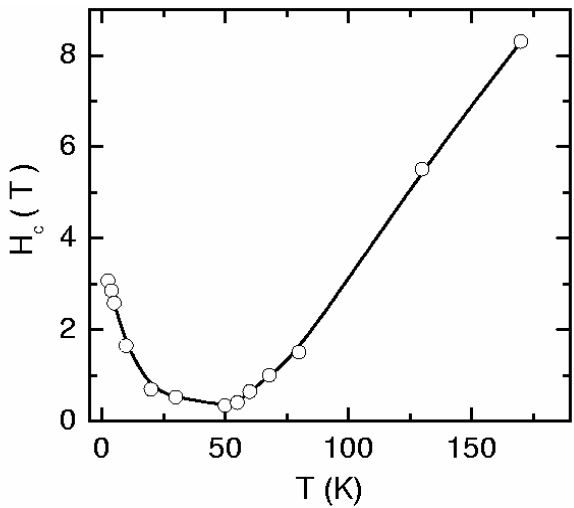

*Fig. 2: The $H_c$ of $Nd_{0.50}Ca_{0.47}Ba_{0.03}MnO_3$ as a function of temperature.*

Figure 1 depicts the magnetization and resistivity as functions of applied magnetic field at 2.5 K. NCBMO shows two successive sharp metamagnetic transitions at ~3 T and ~5 T. Thus, the robustness of CE type charge-ordered state is decreased drastically by 3% Ba doping and as a result, $H_c$ is reduced from ~15 T for NCMO to 3 T for NCBMO. Here, we note that the 3% Ba doping has increased the average $A$-site cation radius ($<r_A>$) from 1.17 Å to 1.18 Å and $\sigma^2$ from negligible ($7\times10^{-5}$ Å$^2$) to 0.0027 Å$^2$, respectively, for NCMO and NCBMO. Furthermore, a fully saturated ferromagnetic state is attained at ~11 T in NCBMO manganite (the saturated magnetization expected for half-doped manganites is 3.5 $\mu_B$ / f.u.). On further magnetic-field cycles, this manganite remains in the ferromagnetic state. This reveals the bi-stable magnetic state of NCBMO in contrast to NCMO which retains its initial antiferromagnetic state after a magnetic field cycle is over. The magnetic history effect vanishes when the NCBMO sample is warmed up to above 250 K. The same sample was used to measure the resistivity in varying magnetic field at 2.5 K (Fig. 1). On increasing the magnetic field, sharp steps in decreasing resistivity are observed, indicating the melting the charge-ordering with electronic transition from insulating to conducting state at 2.5 K. These resistivity-steps correspond very well to the sharp metamagnetic transitions.

Figure 2 displays the values of $H_c$ at different temperatures. Before each measurement of magnetization vs. magnetic field, the sample was warmed up to 270 K to remove the magnetic history. Initially, with increasing temperature, the $H_c$ is decreasing. At 50 K, the $H_c$ passes through a minimum and then keeps increasing with increasing temperature. In case of undoped NCMO, the $H_c$ monotonically decreases with increasing temperature [2]. Thus, the behavior of $H_c$ is contrasting for NCMO and NCBMO manganites. However such minimum in $H_c$ is earlier observed in Ba- and Ce- doped $Pr_{0.50}Ca_{0.50}MnO_3$ system [4,5].

Figure 3 shows the magnetization vs. temperature plots for NCBMO sample. An onset of a magnetic transition is observed below 100 K. A more detailed investigation indicates towards the growth of ferromagnetic clusters in background of antiferromagnetic matrix in this manganite below 100 K. The bifurcation between ZFC and FC magnetization increases very largely with a transition below 41 K. This transition corresponds to an onset of spin glass-like state as observed by the frequency dependent ac susceptibility of NCBMO. Also, such a phase-separated state is earlier observed in similar manganites at low temperatures [4]. However, pure NCMO does not show a phase-separated magnetic state [2]. Therefore, we believe that the origin of minimum of $H_c$ in NCBMO manganite is related to the low tempeature ferromagnetic interactions or a spin glass-like content which is induced by *A*-site cation disorder.

In summary, we have studied the effects of Ba doping on magnetic and electronic properties of half-doped $Nd_{0.50}Ca_{0.47}Ba_{0.03}MnO_3$ manganite. As a result of largely increased cation disorder, we observed successive strongly correlated metamagnetic and electronic transitions. We found that the robustness of CE-type charge-ordered state and hence the $H_c$ drastically decreases by inducing an optimal cation disorder. The $H_c$ of undoped $Nd_{0.50}Ca_{0.50}MnO_3$ manganite decreases with increasing temperature whereas the $H_c$ of $Nd_{0.50}Ca_{0.47}Ba_{0.03}MnO_3$ varies nonmonotonically and passes through a minimum at around 50 K. The origin of this minimum may be related to the glassy-like content or the presence of ferromagnetic interactions in such manganites.

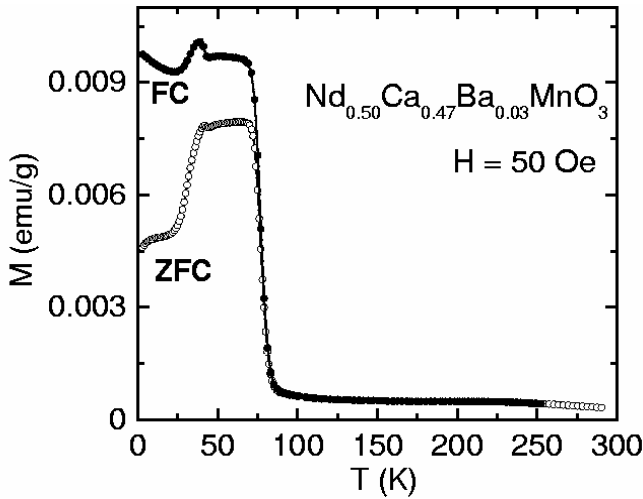

*Fig. 3: Magnetization vs. temperature plots for $Nd_{0.50}Ca_{0.47}Ba_{0.03}MnO_3$ manganite at 50 Oe, in zero-field-cooled (ZFC) and field-cooled (FC) conditions.*

***